\documentclass{jps-cp}
\usepackage{txfonts}
\usepackage{dcolumn}% Align table columns on decimal point
\usepackage{bm,color}% bold math   \textcolor{red} ON
\title{Perturbation Scheme for the Effective Nuclear Force}

\author{Kun Ratha \textsc{Kean}$^{1,2}$, Takashi \textsc{Nishikawa}$^{2}$, and Yoritaka \textsc{Iwata}$^{3}$}

\inst{$^{1}$ Japan Agency of Atomic Energy, Ibaraki, Japan \\
$^{2}$Institute of Innovative Research, Tokyo Institute of Technology, Tokyo, Japan \\
{
$^{3}$Kansai University, Osaka, Japan}
}

\email{iwata{\_}phys@08.alumni.u-tokyo.ac.jp}

\recdate{July 14, 2019}

\abst{
Based on a systematic Hartree-Fock+BCS calculation, new perturbation scheme is proposed to modify the existing Skyrme-type effective nuclear force.
Much attention is paid to have a better description of fission property of heavy nuclei.
In terms of incorporating the nuclear medium effects on the fission barrier height more correctly, two typical Skyrme parameter sets (SkM* and SLy4) are modified.
In conclusion, according to the comparison to experiments, the calculated fission barrier heights are improved by almost 90$\%$.  

}

\kword{Hartree-Fock+BCS, nuclear Skyrme interaction, new perturbation scheme}

\begin{document}
\maketitle
  
\section{Constrained Hartree-Fock+BCS theory}
%\subsection{Microscopic calcultion for many-nucleon systems}
For carrying out the density functional calculation, the constrained Hartree-Fock+BCS theory (CHF+BCS) is utilized to impose a constraint on the quadrupole deformation. 
The master equation is obtained by the variational principle:
\[
\delta \left< \psi | {\mathcal H}  - \beta Q | \psi \right> = 0,
\]
where ${\mathcal H} $ means the Hamiltonian operator of many nucleon systems, the quadrupole parameter $\beta$ plays a role of the Lagrangian multiplier for the quadrupole constraint $\beta Q$, and the trial function $\psi$ is taken as the Slater determinant.
The BCS-type pairing interaction is included in $H$ together with the nuclear and the Coulomb interactions.
Each deformed state and the corresponding energy surface are obtained by choosing the value of $\beta$.

\begin{table}[t] 
\label{table1}
\caption{Skyrme parameter sets (10 parameters profiling the Skyrme-type effective nuclear force; for the detail, e.g., see \cite{98chabanat}).    \\}
\begin{center} 
  \begin{tabular}{l||r|r||r|r}  \hline \hline
                & SLy4 \quad &  SkM* \quad & mSLy4 \quad & mSkM*  \quad  \\ \hline
$t_0$~(MeV$\cdot$fm$^3$) & $-2488.913$ & $-2645.000$ & $-2488.913$ & $-2645.000$  \\
$t_1$~(MeV$\cdot$fm$^5$) & \quad 486.818 & \quad 410.000 & \quad 486.818 & \quad 410.000 \\
$t_2$~(MeV$\cdot$fm$^5$) & ~$-546.395$ & ~$-135.000$ & ~$-546.395$ & ~$-135.000$ \\
$t_3$~(MeV$\cdot$fm$^{3(1+\alpha)}$) & 13777.000 & 15595.000 & 14645.710 & 16004.450 \\
$x_0$ &  0.834 &0.090 &  0.834 &0.090 \\
$x_1$ & $-0.344$ & 0.000 & $-0.344$ & 0.000 \\
$x_2$ & $-1.000$ & 0.000 & $-1.000$ & 0.000 \\
$x_3$ &   $1.354$ & 0.000 &   $1.354$ & 0.000 \\
$W_0$~(MeV$\cdot$fm$^5$) & \quad $123.000$ &\quad 130.000 & \quad $123.000$ &\quad 130.000 \\
$\alpha$ &\quad 0.166667 & \quad 0.166667  &\quad 0.195967 & \quad 0.179000 \\
 \hline \hline
\end{tabular} 
\end{center}
\end{table}

\begin{table*}[t]
\label{table3} 
\caption{Binding energy per nucleon [MeV].
 The calculated values are compared to the experimental value.   \\}
\begin{center}
  \begin{tabular}{c||c||c|c||c|c}  \hline \hline
                & Experiment\cite{nudat2} &  SLy4 &  SkM*  & mSLy4  & mSkM*    \\ \hline
 $^{208}$Pb &$7.87$  & $7.87$& 7.85 & $7.86$    & $7.80$  \\
 $^{120}$Sn &$8.50$  & $8.50$ & 8.48 & $8.52$  & $8.45$ \\
 $^{90}$Zr &$8.71$  & $8.73$ & 8.48  & $8.80$    & $8.45$  \\
 $^{40}$Ca &$8.55$  & $8.66$& 8.63 &  $8.81$   & $8.67$ \\
 $^{16}$O &$7.98$ &  8.10 & 8.14  &  8.37  & 8.25   \\ 
 \hline \hline
\end{tabular} 
\end{center}
\end{table*}

\begin{table*}[t] 
 \label{table4}
\caption{Nuclear radius (proton radius - neutron radius - total radius) [fm].  The calculated values are compared to the experimental value (proton radius). \\
}
\begin{center} 
  \begin{tabular}{c||c||c|c||c|cc}  \hline \hline
                &  Experiment\cite{iaea} &  SLy4  & SkM* &  mSLy4  & mSkM*      \\ \hline
 $^{208}$Pb &$5.50$  & 5.46~-~5.62~-~5.55 & 5.53~-~5.62~-~5.56 & 5.70~-~5.84~-~5.78  & 5.58~-~5.74~-~5.68 \\
 $^{120}$Sn &$4.65$  & 4.59~-~4.73~-~4.68 & 4.58~-~4.73~-~4.67  & 4.79~-~4.92~-~4.86 & 4.68~-~4.83~-~4.77 \\
 $^{90}$Zr & $4.27$  & 4.23~-~4.28~-~4.26  & 4.23~-~4.28~-~4.26 & 4.40~-~4.47~-~4.43 & 4.32~-~4.37~-~4.35\\
 $^{40}$Ca &$3.48$  & 3.45~-~3.38~-~3.41 & 3.46~-~3.40~-~3.43 & 3.57~-~3.50~-~3.53 & 3.53~-~3.47~-~3.50 \\
 $^{16}$O &$2.70$ & 2.79~-~2.72~-~2.76 & 2.77~-~2.71~-~2.74 & 2.83~-~2.76~-~2.80 &2.84~-~2.78~-~2.81  \\ 
 \hline \hline 
\end{tabular} 
\end{center}
\end{table*}

\begin{table}[t]
\label{table2}
\caption
{
Nuclear matter properties calculated by \cite{jirina} (for the detail, see~\cite{12dutra}).   
From left side, energy per nucleon $E/A$, imcompressivity $K$, symmetry energy $S$, slope of the symmetry energy $L$ are shown.
}
\begin{center}  \begin{tabular}{l|ccccc}
\hline \hline
parameter & $E/A$[MeV] & $K$[MeV]  & $S$[MeV] & $L$[MeV] \\ \hline
SkM*      & -15.77   & 216          & 30     & 46    \\
SLy4      & -15.97   & 231         & 32     & 46     \\  \hline
mSkM*     & -15.39   & 234          & 30     & 44     \\
mSLy4     & -15.21   & 269         & 31     & 32     \\
 \hline \hline
\end{tabular} \end{center}
\end{table}

%\subsection{Numerical calculations}
The calculation is performed using the SkyAX code \cite{99reinhard}  in which the quadrupole deformation is given on the three-dimensional Cartesian coordinate.
In the SkyAX code, the octupole moment is optimized by adding a small octupole moment to the initial wave functions under the quadrupole constraint.
Although the axial symmetry is assumed for the SkyAX calculations, it does not require anything more for the quadrupole constraint calculations.
Indeed, the quadrupole-deformed nuclei can be fully described within the axial symmetric framework.

\begin{table}[htb] \label{table0}
\caption{Fission barrier height of heavy nuclei [MeV]: inner barrier (left) and outer barrier (right). }
\begin{center}
  \begin{tabular}{c||c|c|c||c|c}  \hline \hline
                &  Experiment (RIPL-2~\cite{ripl2}) &  SLy4  & SkM*  &  mSLy4  & mSkM*        \\ \hline 
                
% $^{230}$Th &$6.10,~ 6.80$  & $7.21,~ 9.57$ & $5.73,~ 6.63$ & $4.12,~ 6.17$ & $3.80 ,~ 4.75$  \\ 
% $^{231}$Th &$6.00,~ 6.70$  & $7.36,~ 9.12$ & $5.91,~ 6.35$ & $4.27,~ 6.17$ & $4.03 ,~ 4.80$  \\
% $^{232}$Th &$5.80,~ 6.70$  & $7.58,~ 9.05$ & $6.25,~ 6.28$ & $4.50,~ 6.42$ & $4.32 ,~ 4.95$  \\
 
% $^{230}$Pa &$5.60,~ 5.80$  & $7.07,~ 9.65$ & $5.56,~ 6.28$ & $3.82,~ 5.43$ & $3.54,~ 4.15$   \\ 
% $^{231}$Pa &$5.50,~ 5.50$  & $7.14,~ 8.71$ & $5.65,~ 6.05$ & $4.01,~ 5.47$ & $3.74,~ 4.11$   \\
% $^{232}$Pa &$5.00,~ 6.40$  & $7.46,~ 8.61$ & $6.04,~ 6.03$ & $4.19,~ 5.65$ & $4.03,~ 4.19$   \\
% $^{233}$Pa &$5.70,~ 5.80$  & $7.92,~ 8.80$ & $6.39,~ 5.97$ & $5.86,~ 4.44$ & $4.34,~ 4.35$   \\
% $^{234}$Pa &$6.30,~ 6.15$  & $8.57,~ 9.18$ & $6.76,~ 6.08$ & $4.70,~ 6.11$ & $4.66,~ 4.55$   \\

 $^{231}$U &$4.40,~ 5.50$  & $6.90,~ 6.51$ & $5.34,~ 5.50$ & $3.63,~ 4.72$ & $3.36,~ 3.37$    \\
 $^{232}$U &$4.90,~ 5.40$  & $7.10,~ 8.66$ & $5.63,~ 5.48$ & $3.84,~ 4.84$ & $3.65,~ 3.45$    \\
 $^{233}$U &$4.35,~ 5.55$  & $7.49,~ 8.09$ & $6.01,~ 5.55$ & $4.11,~ 4.99$ & $4.01,~ 3.56$    \\
 $^{234}$U &$4.80,~ 5.50$  & $8.02,~ 8.31$ & $6.45,~ 6.60$ & $4.38,~ 5.28$ & $4.35,~ 3.76$    \\
 $^{235}$U &$5.25,~ 6.00$  & $8.77,~ 8.75$ & $6.99,~ 5.86$ & $4.68,~ 5.56$ & $4.74,~ 3.97$    \\  
 $^{236}$U &$5.00,~ 5.67$  & $7.36,~ 8.29$ & $7.53,~ 6.18$ & $5.00,~ 5.89$ & $5.09,~ 4.27$    \\
 $^{237}$U &$6.40,~ 6.15$  & $9.86,~ 8.52$ & $7.93,~ 6.46$ & $5.30,~ 6.15$ & $5.37,~ 4.50$  \\     
 $^{238}$U &$6.30,~ 5.50$  & $10.2,~ 8.92$ & $6.37,~ 6.97$ & $5.60,~ 6.51$ & $5.66,~ 4.78$    \\
 $^{239}$U &$6.45,~ 6.00$  & $10.7,~ 9.28$ & $8.70,~ 8.21$ & $5.88,~ 6.90$ & $5.88,~ 5.05$    \\

 $^{236}$Np &$5.90,~ 5.40$ & $8.87,~ 8.27$ & $7.11,~ 5.49$ & $4.63,~ 4.89$ & $4.79,~ 3.40$    \\
 $^{237}$Np &$6.00,~ 5.40$ & $9.52,~ 8.39$ & $7.66,~ 7.76$ & $4.98,~ 5.21$ & $5.15,~ 3.67$    \\
 $^{238}$Np &$6.50,~ 5.75$ & $10.0,~ 8.35$ & $8.16,~ 6.18$ & $5.29,~ 5.51$ & $5.48,~ 3.93$    \\

 $^{237}$Pu &$5.10,~ 5.15$  & $8.92,~ 7.69$ & $7.17,~ 4.50$ & $4.55,~ 4.17$ & $4.77,~ 2.74$   \\
 $^{238}$Pu &$5.60,~ 5.10$  & $9.55,~ 8.14$ & $7.75,~ 5.34$ & $4.89,~ 4.47$ & $5.16,~ 2.98$   \\
 $^{239}$Pu &$6.20,~ 5.70$  & $10.1,~ 8.67$ & $8.25,~ 5.62$ & $5.25,~ 4.85$ & $5.55,~ 3.31$   \\
 $^{240}$Pu &$6.05,~ 5.15$  & $10.6,~ 8.79$ & $8.76,~ 6.08$ & $5.59,~ 5.25$ & $5.87,~ 3.62$   \\
 $^{241}$Pu &$6.15,~ 5.50$  & $11.2,~ 8.81$ & $9.19,~ 6.34$ & $5.90,~ 5.64$ & $6.20,~ 3.96$   \\
 $^{242}$Pu &$5.85,~ 5.05$  & $11.6,~ 8.92$ & $9.48,~ 6.69$ & $6.20,~ 6.06$ & $6.39,~ 4.21$   \\
 $^{243}$Pu &$6.05,~ 5.45$  & $11.8,~ 9.07$ & $9.56,~ 6.97$ & $6.46,~ 6.47$ & $6.52,~ 4.41$   \\
 $^{244}$Pu &$5.70,~ 4.85$  & $11.9,~ 9.34$ & $9.49,~ 7.19$ & $6.65,~ 6.60$ & $6.59,~ 4.50$   \\
 $^{245}$Pu &$5.85,~ 5.25$  & $11.7,~ 10.2$ & $9.37,~ 7.52$ & $6.73,~ 6.43$ & $6.53,~ 4.16$   \\
 
 $^{239}$Am &$6.00,~ 5.40$  & $9.51,~ 7.50$ & $7.74,~ 4.75$ & $4.80,~ 3.74$ & $5.15,~ 2.29$   \\
 $^{240}$Am &$6.10,~ 6.00$  & $10.1,~ 8.00$ & $8.29,~ 5.05$ & $5.19,~ 4.15$ & $5.59,~ 2.66$   \\ 
 $^{241}$Am &$6.00,~ 5.35$  & $10.7,~ 8.66$ & $8.87,~ 5.43$ & $5.56,~ 4.53$ & $5.96,~ 2.99$   \\
 $^{242}$Am &$6.32,~ 5.78$  & $11.3,~ 8.54$ & $9.35,~ 5.89$ & $5.90,~ 4.97$ & $6.27,~ 3.29$   \\
 $^{243}$Am &$6.40,~ 5.05$  & $11.9,~ 8.57$ & $9.74,~ 6.26$ & $6.22,~ 5.35$ & $6.52,~ 3.59$   \\
 $^{244}$Am &$6.25,~ 5.90$  & $12.1,~ 8.74$ & $9.81,~ 8.52$ & $6.49,~ 5.75$ & $6.68,~ 3.78$   \\

 $^{241}$Cm &$7.15,~ 5.50$  & $10.1,~ 7.39$ & $8.35,~ 4.45$ & $5.12,~ 3.39$ & $5.56,~ 1.92$   \\
 $^{242}$Cm &$6.65,~ 5.00$  & $10.7,~ 8.04$ & $9.00,~ 4.94$ & $5.49,~ 3.79$ & $5.96,~ 2.28$   \\  
 $^{243}$Cm &$6.33,~ 5.40$  & $11.5,~ 8.25$ & $9.53,~ 5.32$ & $5.84,~ 4.21$ & $6.30,~ 2.61$   \\ 
 $^{244}$Cm &$6.18,~ 5.10$  & $12.0,~ 8.21$ & $9.91,~ 9.74$ & $6.18,~ 4.59$ & $6.54,~ 2.87$   \\
 $^{245}$Cm &$6.35,~ 5.45$  & $12.3,~ 8.51$ & $10.0,~ 6.02$ & $6.46,~ 5.02$ & $6.76,~ 3.09$   \\
 $^{246}$Cm &$6.00,~ 4.80$  & $12.4,~ 9.14$ & $10.0,~ 6.25$ & $6.68,~ 4.91$ & $6.83,~ 3.12$   \\
 $^{247}$Cm &$6.12,~ 5.10$  & $12.2,~ 9.86$ & $9.92,~ 6.58$ & $6.77,~ 4.63$ & $6.79,~ 2.85$   \\
 $^{248}$Cm &$5.80,~ 4.80$  & $11.8,~ 10.7$ & $9.68,~ 6.02$ & $6.77,~ 4.23$ & $6.72, ~2.52$   \\              
 $^{249}$Cm &$5.63,~ 4.95$  & $11.5,~ 10.4$ & $9.40,~ 5.50$ & $6.65,~ 3.79$ & $6.56,~ 2.14$   \\
 \hline \hline
\end{tabular} 
\end{center}
\end{table}

\section{Doubly-constrained Skyrme perturbation scheme}
Let $\rho$ be the density.
For Skyrme-type effective nuclear interaction \cite{72vautherin}, the Hamiltonian density reads
\begin{equation} \label{eqterms}
 H  = t_0 \rho^2  + \frac{t_3}{6} \rho^{2+\alpha} + \cdots.
\end{equation}
We focus on the competition between $t_0 \rho^2$ (two-body force) and $ t_3 \rho^{2+\alpha}$ (medium effect) to have a better description of fission and a better prediction power for heavy nuclear physics in general. 
The perturbation scheme optimizes the nuclear medium effects (medium effect); more precisely, nuclear density-dependent term ($t_3$-term) of the Skyrme interaction.
The refit protocol of Skyrme perturbation is shown as follows: \\
%\begin{itemize}
%\item 
first adding a perturbation $\delta \alpha$ to fit the theoretically calculated fission barrier height to the experiment; 
%\item
second changing the parameter $t_3$ to keep the quality of the original interaction in terms of reproducing the static properties; 
%\end{itemize}
A whole $t_3$-term is optimized as $t_3' \rho^{2+\alpha + \delta \alpha}$ to reproduce the experimental fission property, where $\delta a$ and $t_3'$ are the perturbation and the modified-value for $t_3$, respectively.
One of the two reference quantities is the fission barrier height  (to be compared to \cite{ripl2}), and the other is the binding energy (to be compared to \cite{nudat2}).
The Skyrme perturbation scheme is judged to be working correctly only if
\begin{equation} \label{cond1}
 \left| 1- \frac{  <\psi| t_3' \rho^{2+\alpha + \delta \alpha}| \psi> }{ <\psi| t_3 \rho^{2+\alpha}| \psi> } \right| < 0.05
\end{equation}
is satisfied.
%%%
This condition of permitting 5$\%$ difference is rather reasonable, because almost a few $\%$ differences in binding energy already exist in the original interactions.
The proposed Skyrme perturbation theory makes use of the different mathematical behavior due to the power index ($\alpha$) and the multiplication parameter ($t_3$).

\section{Results}
Two Skyrme-type effective nuclear interactions mSkM* and mSLy4 interactions are proposed by modifying SkM* \cite{82bartel} and SLy4 \cite{98chabanat}, respectively
The differences between the existing theoretical values and the experimental values (5.00~MeV~\cite{ripl2}) are $2.36$~MeV for SLy4 and $2.53$~MeV for SkM*. 
The experimental fission barrier height for $^{236}$U (inner barrier) is utilized to modify the effective nuclear force parameters, and  the binding energy of $^{236}$U is also fitted to keep the original quality of static calculations.

The obtained interaction is shown in Table I.
For the validity check, the binding energies and the radii of $^{16}$O, $^{40}$Ca, $^{90}$Zr, $^{120}$Sn, and $^{208}$Pb are also utilized (Table II and III), as they are referred to in most of Skyrme parameter fittings. 
For the quality check, nuclear matter properties are compared in Table IV.
The fission barrier height is systematically calculated for Uranium to Curium isotopes (Table V).
Calculated isotopes are selected based on whether the comparable experimental data \cite{ripl2} exist or not.
Possibly due to the inclusion of the pairing interaction, the quality of barrier height calculations for odd-odd, odd-even and even-odd nuclei are as good as those for even-even nuclei, where odd nuclei are well calculated without blocking.

\section{Discussion}
The Skyrme-type effective nuclear forces typically overestimate the fission barrier heights, which prevents us to have a precise prediction/control of nuclear fission processes.
The proposed interaction shows a significant improvement of the calculated fission barrier height not only for Uranium isotopes but also for the other isotopes.
As an average, for both two interactions, the description of fission barrier height is improved by almost 90$\%$.
The proposed interactions somewhat underestimate the saturation density, which will be addressed in the future research.
%%%
%For the nuclear matter properties, we see that the difference between existing and proposed interactions are no larger than the order of the difference between the existing interactions (for example, see \cite{09klupfel}) except for the baryonic pressure.

\end{document}